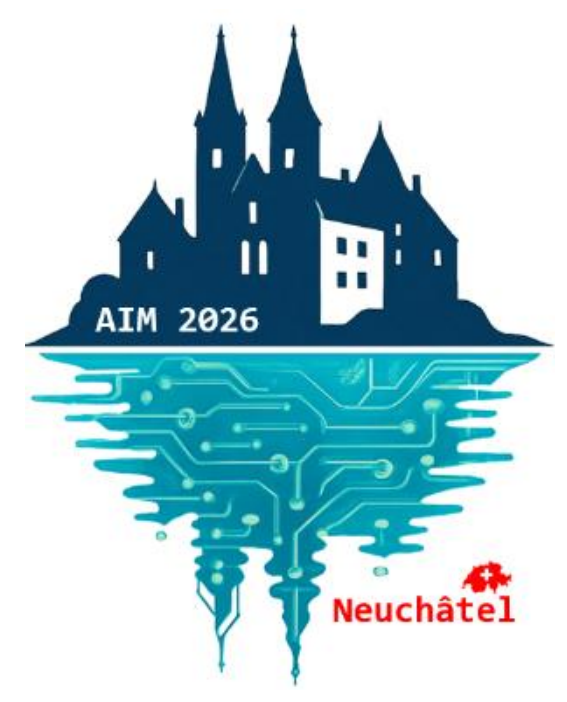

---

# Derrière le Contenu : Vues Mobiles de Wikipédia et Activité Touristique

**Lucas Eustache**, University Paris-Dauphine
**Paul Favier**, University Paris-Dauphine

**Résumé**

Cette étude examine si les traces numériques ouvertes peuvent fournir des indicateurs interprétables et à haute fréquence de l'activité touristique locale. Nous soutenons que la composition par appareil de l'attention portée à Wikipédia permet de distinguer l'usage informationnel situé de la planification à distance : les vues mobiles reflètent davantage des besoins informationnels contemporains et sur site, tandis que les vues desktop capturent un intérêt plus diffus dans le temps. En reliant les nuitées hôtelières quotidiennes d'Accor au trafic des pages Wikipédia de 704 communes françaises entre 2018 et 2025, nous montrons que les vues mobiles sont positivement associées à la demande hôtelière du même jour et dominent le trafic desktop dans les spécifications conjointes. Cette relation est plus forte dans les destinations orientées loisirs et dans les lieux à forte visibilité Wikipédia. Une validation microéconomique sur six attractions culturelles d'Orléans confirme ce mécanisme. Les vues mobiles de Wikipédia constituent ainsi un signal ouvert et réplicable de *nowcasting* touristique.

# Behind the Content: Wikipedia Mobile Views and Tourism Activity

**Abstract**

This study examines whether open digital traces can provide interpretable, high-frequency indicators of local tourism activity. We argue that the device composition of Wikipedia attention helps distinguish situated information use from remote planning: mobile pageviews are more likely to reflect on-site, contemporaneous information needs, whereas desktop pageviews capture temporally diffuse interest. Linking daily Accor hotel room-nights to Wikipedia city-page traffic for 704 French communes from 2018 to 2025, we find that mobile pageviews are positively associated with same-day hotel demand and dominate desktop traffic in joint specifications. The relationship is stronger in leisure-oriented destinations and in places with higher Wikipedia visibility. A micro-validation using daily attendance at six cultural attractions in Orléans shows the same pattern: mobile pageviews predict same-day gate counts, while surrounding leads and lags are close to zero. The findings position mobile Wikipedia traffic as a transparent, replicable nowcasting signal for tourism activity.

# Behind the Content: Wikipedia Mobile Views and Tourism Activity

## 1. Introduction

Organizations increasingly use digital trace data to infer demand, monitor behaviour, and allocate resources in near real time (Choi & Varian, 2012; Bańbura et al., 2013). A persistent challenge is interpretability: the same online signal can reflect heterogeneous intentions (curiosity, learning, planning, or action), and its mapping to offline outcomes is often ambiguous. This challenge is particularly salient when managers and policymakers seek high-frequency indicators of local economic activity, but the most precise measures (e.g., mobile location data, transaction records) are proprietary and costly. The question is whether behaviourally meaningful structure can be extracted from open digital traces?

Tourism provides a canonical setting where interpretability is both difficult and consequential. Travel decisions are experience-based; tourists face substantial information frictions and rely on ongoing search before and during trips (Buhalis & Law, 2008; Xiang et al.,2015). Cities, as policy-relevant units, often lack accessible indicators of visitation: traditional tourism statistics are released with delays, while proprietary alternatives are costly and unavailable to most stakeholders. Episodic shocks, from transport disruptions to public health restrictions, further underscore the need for transparent, replicable measurement tools. Research on search-engine data has shown that online query volumes can improve tourism demand forecasts (Bangwayo-Skeete and Skeete, 2015; Yang et al., 2015; Li et al., 2021), yet we know less about whether the composition of online information consumption contains incremental signal about contemporaneous, on-the-ground tourism activity.

This paper argues that where and how information is accessed is relevant to that signal. Mobile devices embed information seeking in time and space: smartphones reduce the friction of "in-the-moment" queries but impose interface constraints that alter browsing costs and depth (Ghose, Goldfarb, et al., 2013). Desktop access is more consistent with remote, leisurely planning. Prior work demonstrates that mobile contexts heighten locality and immediacy; in tourism specifically, smartphones support navigation, re-planning, and coordination during travel (Dickinson et al., 2016; Wang et al., 2016; Huertas & Orden-Mejía, 2022). These mechanisms imply that device-based differences in information consumption can serve as a theoretically grounded proxy for the timing and context of offline activity.

We study this idea using Wikipedia, a general-purpose, non-transactional information infrastructure whose content is publicly accessible and widely consulted. Unlike vertically integrated travel platforms, Wikipedia functions as a reference layer that travelers use to resolve factual uncertainty (Greenstein & Zhu, 2016; Greenstein & Zhu, 2018). A causal field experiment shows that improving city-page content can increase tourism outcomes (Hinnosaar et al., 2023). These features make Wikipedia particularly useful for disentangling information consumption as behaviour from platform-mediated transactions.

Our core measurement insight is that device composition can sharpen the behavioural interpretation of online attention. Total pageviews of a city's Wikipedia page may reflect diffuse salience (e.g., education, news, general curiosity) and thus provide a noisy proxy for visitation. By contrast, the share of mobile pageviews plausibly captures the fraction of attention generated in "on-the-go" contexts that align more tightly with contemporaneous physical presence. This yields our two research questions:

RQ1: Do daily mobile Wikipedia pageviews serve as a high-frequency proxy for same-day hotel room-night demand at the city level?

RQ2: How does the Wikipedia–tourism relationship vary across cities, and what city characteristics explain this heterogeneity?

We answer these questions using a daily city-level panel linking Accor hotel data to Wikipedia mobile pageviews for 704 French cities over January 2018 to August 2025 (approximately 1.6 million city-day observations). Leveraging within-city, within-date variation through two-way fixed effects (Wooldridge, 2010; Angrist & Pischke, 2009), we find an elasticity-like coefficient of +0.125. We validate the physical-presence mechanism with a micro-level analysis of six cultural attractions in Orléans.

This study contributes to the literature in three ways. First, it advances research on digital traces and nowcasting (Choi & Varian, 2012; Bańbura et al., 2013; Giannone et al., 2008) by demonstrating that daily-frequency mobile Wikipedia data can complement nowcasting tourism indicator, extending prior work that relies on monthly or annual aggregates. Second, it addresses the interpretability problem by showing that composition-based measures (mobile versus desktop) improve signal quality through a device-specific behavioural mechanism, contributing to the IS literature on how platform features shape economic inference (Ghose, Goldfarb, et al., 2013; Dellarocas, 2003). Third, it provides one analysis of city-level heterogeneity in the Wikipedia–tourism relationship, showing that this relationship is linked to information needs.

## 2. Literature Review

This paper draws on two research streams: digital traces for measuring economic activity, and device context as a behavioral filter for online information consumption.

## 2.1. From Offline Activity to Online Activity

A broad IS literature establishes that digital traces co-move with offline demand. Online word-of-mouth predicts product sales (Godes & Mayzlin, 2004; Chevalier & Mayzlin, 2006), and the digitization of consumer voice enables observation of demand signals that were previously hidden (Dellarocas, 2003). In travel markets, user-generated content shapes consumer choice and firm strategy: crowd-sourced hotel reviews improve preference matching (Ghose et al., 2012), ratings drive advertising responses (Hollenbeck et al., 2019), and interventions altering review generation produce marketplace effects (Gao et al., 2025). In macroeconomics and tourism forecasting, search-engine query volumes have become a standard nowcasting input (Choi & Varian, 2012; Bangwayo-Skeete & Skeete, 2015; Yang et al., 2015). These studies support the premise that online information use co-moves with economic activity, especially in experience-good settings

Yet the established association is typically with aggregate or forward-looking demand, and the primary discriminant is the level of activity. For real-time local measurement, the question is narrower: among all people consulting information about a destination on a given day, which fraction is physically present? This is an identification problem, because the same Wikipedia page can be read by a student, a family planning a trip, or a tourist at the train station.

Consumer search theory clarifies when information consumption tracks behavior most tightly. Under costly sequential search, timing depends on uncertainty and time pressure (Bakos, 1997; De Los Santos et al., 2012). Tourism is a canonical high-friction setting where information needs evolve from planning to on-site execution (Buhalis and Law, 2008; Dickinson et al., 2016; Huertas and Orden-Mejía, 2022). Search observed during a trip, when uncertainty is immediate, should map more directly onto local presence than search observed weeks earlier. A trace feature that discriminates between these contexts is needed.

## 2.2. Device Composition as a Behavioral Signal

Mobile and desktop access embed information seeking in fundamentally different contexts, providing the discriminating feature. Ghose, et al. (2013) demonstrate that mobile usage across physical locations increases the salience of local activities, implying that mobile queries are more situated and locally oriented than desktop queries. In tourism, on-site tourists use smartphones for navigation, logistics, and context-specific queries (Dickinson et al., 2016; Wang et al., 2016; Huertas & Orden-Mejía, 2022), while desktop browsing aligns with at-home planning. Device type thus acts as a behavioral filter: it separates situated, contemporaneous information use from temporally diffuse, remote browsing.

Applied to Wikipedia pageviews, this distinction generates a concrete measurement insight. Total pageviews for a city page may reflect diffuse interest and thus provide a noisy proxy. The mobile page view plausibly captures the fraction of readers consuming the page while physically traveling. Because the measure is a composition, it can contain signal beyond levels: holding total attention constant, a shift from desktop to mobile indicates a transition from low-intent to situated consumption that should map more tightly onto same-day presence.

Wikipedia is particularly clean for testing this mechanism. Unlike vertically integrated travel marketplaces, Wikipedia is a general-purpose, non-transactional information infrastructure oriented toward encyclopedic coverage (Greenstein & Zhu, 2016). Mobile Wikipedia use is therefore unlikely to be driven by platform-specific nudges (push notifications, conversion-optimized interfaces) that could confound the behavioral interpretation. Crucially, a randomized field experiment shows that enriching tourist-relevant Wikipedia content increases overnight stays through increased readership (Hinnosaar et al., 2023), confirming that Wikipedia functions as an active intermediary in the tourism information chain. Applied work has further explored Wikipedia traffic as a proxy for visitation patterns (Owuor et al., 2023), though prior studies use monthly or annual aggregates and do not exploit the device-composition mechanism we propose.

### 2.3. Hypotheses

Hypothesis 1: Mobile pageviews reflect physical presence. When tourists are on site, they generate mobile Wikipedia pageviews because of needs. Mobile pageviews for a city's Wikipedia page should therefore move contemporaneously with actual tourist presence.

Hypothesis 2: Physical presence generates contemporaneous hotel demand. Tourist visits produce hotel room nights. We hypothesize that physical presence is the common cause of both signals, simultaneously driving mobile browsing and accommodation demand. If so, mobile pageviews and room-nights should co-vary within the same city on the same day: $\beta_{mobile} > 0$ in a regression of log room-nights on log mobile pageviews with city and date fixed effects.

Hypothesis 3: Mobile effect exceeds desktop effect. Because mobile pageviews are tightly synchronized with physical presence while desktop pageviews capture more temporally diffuse planning behavior, we expect: $\beta_{mobile} > \beta_{desktop}$ at the daily level.

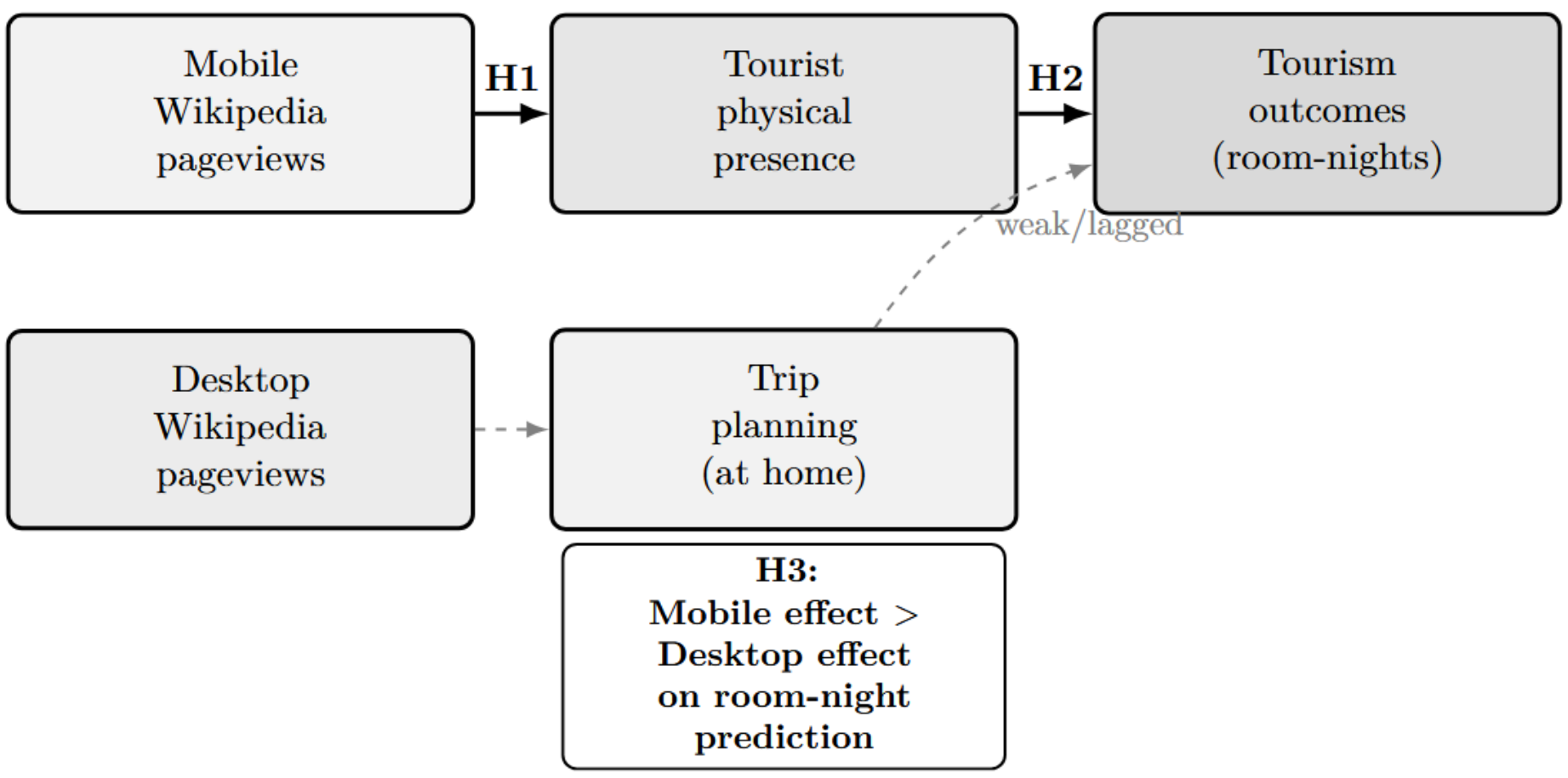

***Figure 1: Conceptual Model and Behavioural Mechanism***

# 3. Data

Our empirical analysis draws on four complementary data sources organised at two levels of geographic resolution. At the national level, we combine daily hotel records from Accor Group with daily Wikipedia city-page pageviews for 704 French communes. At the city level, we match daily gate counts from six cultural attractions in Orléans to their site-specific Wikipedia pageviews. We further incorporate two supplementary INSEE datasets to know the touristic capacity of city by type of accommodation (*camping, hotel ...*). Table 1 presents the construction steps of the national panel.

| Step | Description | Observations | Communes |
|---|---|---|---|
| 1b | Accor raw (reservation records, all channels) | 5,900,336 | 704 |
| 2 | Accor: pivot from long to city-day format | 1,680,331 | 704 |
| 1a | Wikipedia raw (daily city pageviews, Wikimedia API) | 1,927,552 | 704 |
| 3 | Inner join Wikipedia × Accor on (INSEE code, date) | 1,635,454 | 704 |

***Table 1: Data construction pipeline: from raw sources to analytical sample***

## 3.1. National Level: Hotel Data

The hotel-performance data are provided by Accor Group under a proprietary data-sharing agreement. The raw file records 5,900,336 reservation lines covering 704 unique INSEE codes, spanning January 2018 to August 2025, disaggregated by six distribution channels. We pivot to a city-day format, summing room-nights across all channels to obtain $ROOM_NIGHT_TOTAL_{it}$. Our primary dependent variable is:

$$\text{LogRN}_{it} = \log\left(1 + \text{ROOM_NIGHT_TOTAL}_{it}\right).$$

The raw variable is strongly right skewed: with a mean of 93.2 and a median of 35.0 (SD: 434.9), the distribution is dominated by large urban destinations. The log transformation reduces skewness from 23.9 to 0.29, enabling an elasticity like coefficient interpretation and stabilising the variance (Wooldridge, 2010).

## 3.2. National Level: Wikipedia City Pageviews

For each city, we retrieve daily pageview counts from the Wikimedia API, disaggregated by platform (desktop, mobile-web, mobile-app) and agent type (human, automated). Our primary attention variable uses mobile-web pageviews by human users exclusively; mobile-app traffic is excluded because its daily volume is negligible (mean: 3.3 views per city-day versus 109.7 for mobile-web). We construct:

$$\begin{aligned} \text{MobileLog}_{it} &= \log(1 + \text{MobileWebViews}_{it}), \\ \text{DesktopLog}_{it} &= \log\left(1 + \text{DesktopViews}_{it}\right), \end{aligned}$$

The mean daily mobile-web views is 109.7 (median: 54; SD: 193.2).

### 3.3. City Level: Orléans Cultural Attractions

To complement the national panel with a more direct test of the physical-presence mechanism, we use a micro-level dataset from Orléans (Loiret, France)[1]. Daily entry counts for six cultural attractions are recorded at the gate level over 1,237 consecutive days (1 July 2021 to 18 November 2024), yielding 7,422 place-day observations. Unlike hotel room-nights, gate counts are a direct measure of physical presence rather than a proxy through accommodation demand.

The six sites span a wide range of visitor volumes, from the *Hôtel Groslot* and *Parc floral de la Source* (over 255 mean daily visitors) to the *Musée historique et archéologique* (16 mean daily visitors). The 14.5% rate of zero-attendance days, reflecting closure days and seasonal shutdowns, enriches identification by contrasting attended and non-attended periods. For each of the six Orléans attractions, we retrieve daily pageviews from the Wikimedia API for the site-specific Wikipedia page (e.g. *Parc floral de la Source*, *Maison de Jeanne d'Arc*), disaggregated by platform. Mean daily mobile pageviews range from 2.6 (Hôtel Groslot) to 5.9 (*Parc floral de la Source*).

## 4. Method

### 4.1. National Level Estimate

We estimate whether Wikipedia traffic contains a contemporaneous signal for tourism demand using two complementary panel designs. The first is a national daily city panel linking hotel room-nights to Wikipedia pageviews at the city-day level. The second is a site-level panel for six cultural attractions in Orléans, where daily attendance is matched to pageviews for the corresponding Wikipedia pages. The empirical strategy is therefore based on within-unit changes over time, after absorbing common shocks with date fixed effects. In the national panel, the baseline specification is a two-way fixed-effects model of the form

$$\log(1 + \mathrm{RN})_{it} = \beta \cdot \mathrm{MobileLog}_{it} + \alpha_i + \lambda_t + \varepsilon_{it},$$

where $\mathrm{RN}_{it}$ denotes hotel room-nights in city $i$ on day $t$, $\alpha_i$are city fixed effects, and $\lambda_t$are date fixed effects. City fixed effects absorb time-invariant differences across cities, such as baseline tourism appeal or accommodation capacity, while date fixed effects absorb shocks common to all cities on a given day, including seasonality, day-of-week effects, and nationwide demand fluctuations. s. Standard errors are computed using the Driscoll–Kraay estimator with a 30-day bandwidth, indexed by calendar date, to allow for cross-sectional dependence across cities and serial correlation over time. We estimate four variants corresponding to the reported national results: a mobile-only model, a desktop-only model, a joint model including both platforms, and an interaction model allowing the coefficients to vary with a city-level tourism-orientation proxy index $P2$. This index is defined as the share of leisure beds in total accommodation capacity (camping bed and holidays camp bed); higher values therefore proxy a more leisure-oriented tourism structure. The main effect of $P2_i$ is absorbed by city fixed effects because $P2_i$ is time-invariant The interaction model can be written as:

[1] Data accessible here and code for replication available here

$\log(1 + \text{RN})_{it} = \beta_1 \cdot \text{MobileLog}_{it} + \beta_2 \cdot \text{DesktopLog}_{it} + \ \beta_3 \cdot (\text{MobileLog}_{it} X\ P2it)\ \alpha_i + \lambda_t + \varepsilon_{it},$

### 4.2. City Level Estimate

The Orléans analysis removes city-level aggregation by matching each attraction's daily attendance directly to its own Wikipedia page. Because both attendance and daily pageviews are low-count and right-skewed mean daily mobile pageviews are 4.2 with a median of 3 all variables enter in the $\log(1 + x)$ form, which retains zero observations while yielding elasticity-like coefficients for positive counts. The outcome is $y_{it} = \log(1 + \text{Attendance}_{it})$ ; the attention variables are $m_{it} = \log(1 + \text{Mobile}_{it})$ and $d_{it} = \log(1 + \text{Desktop}_{it})$. Table 2 reports three contemporaneous specifications: mobile only (column 1), desktop only (column 2), and both device jointly (column 3). All specifications include venue fixed effects, exact-date fixed effects, venue-by-year-month fixed effects, venue-by-day-of-week fixed effects, and one lag of log attendance to control for persistence to account for seasonality and autocorrelation. Standard errors use the Driscoll-Kraay estimator with a 30-day bandwidth indexed by calendar date, which is robust to cross-sectional dependence and serial correlation up to the specified lag window. To reduces concerns about reverse causality, we estimate a distributed lead-lag specification extending the pageview window to $k \in [-30, +30]$ days:

$$\text{Attendance}_{it} = \sum_{k=-30}^{30} \delta_k \cdot \text{MobilePV}_{i,t+k} + \mu_i + \lambda_t + \varepsilon_{it}.$$

A profile concentrated at $k = 0$ with near-zero coefficients at surrounding lags identifies same-day co-movement and is inconsistent with slow-moving attention trends or reverse causality. Because 61 timing coefficients are tested simultaneously, we apply a Bonferroni (Dunn, 2012) correction to control the familywise error rate.

Finally, we assess predictive value by comparing four nowcasting models estimated on the pre-holdout sample and evaluated on the last 180 calendar days of the panel (23 May 2024 onward, N = 1,080 site-days). The baseline model includes venue fixed effects and calendar controls (day-of-week, month, and venue-specific linear trends) but no Wikipedia variables. Then one model augments the baseline with same-day log mobile pageviews, another adds same-day log desktop pageviews, and a last includes both. Forecasting accuracy is measured by RMSE on the attendance scale, mean absolute error, log-scale RMSE, and Pearson correlation. A six-fold expanding-window (one fold per month) monthly support cross validation by assessing predictive power of our model outside of the training set.

## 5. Results

We proceed in two steps. We first establish the aggregate national effect, and we show that this effect is heterogenous, driven by touristic characteristic of the city. Then, we zoom in on six Orléans cultural attractions to validate the mechanism and show when the signal adds the most real-time value, and we additionally show that the effect hold as long as the page has enough popularity, to distinguish the noise from real attention.

### 5.1. National Level

The national estimates show a clear positive association between mobile Wikipedia traffic and hotel demand. In the mobile-only specification, the coefficient on mobile pageviews is 0.1254 and precisely estimated. Interpreted in percentage terms, this implies that a 10% increase in mobile Wikipedia traffic within a city-day is associated with approximately a 1.25% increase in hotel room-nights on that same day. The corresponding desktop-only coefficient is smaller, at 0.0675, indicating that desktop traffic also co-moves with hotel demand but substantially less strongly than mobile traffic.

The joint specification sharpens this contrast. When mobile and desktop pageviews are entered simultaneously, the mobile coefficient remains large at 0.1163, whereas the desktop coefficient falls to 0.0347. Consistent H1 and H3 this attenuation indicates that much of the predictive content in desktop traffic overlaps with the signal already captured by mobile traffic, while mobile traffic retains a distinct and economically meaningful association with same-day hotel demand.

The interaction model further shows that this relationship varies systematically with the tourism structure of the city. The coefficient on mobile pageviews is 0.1411, and the interaction between mobile traffic and $P2$ is 0.1162. The positive interaction implies that the mobile-demand relationship becomes stronger as the local accommodation structure becomes more leisure-oriented. By contrast, the desktop coefficient is 0.0148, while the interaction between desktop traffic and $P2$ is negative ($-0.0229$). Taken together, the national results indicate that mobile Wikipedia traffic is the dominant digital correlate of contemporaneous hotel demand, and that its relevance is greatest in places with a stronger leisure tourism profile.

The reported within-$R^2$ values are modest, ranging from 0.0018 to 0.0140. This is not surprising in a specification with both city and date fixed effects, where identification relies only on residual within-city, within-day variation after common shocks and persistent local differences have already been removed.

| **Terme** | **M1** | **M2** | **M3** | **M4** |
|---|---|---|---|---|
| **Mobile** | **0.1254***<br>*(0.0088)* | — | **0.1163***<br>*(0.0082)* | **0.1411***<br>*(0.0080)* |
| **Desktop** | — | **0.0675***<br>*(0.0061)* | **0.0347***<br>*(0.0043)* | **0.0148**<br>*(0.0054)* |
| **Mobile** x **P2** | — | — | — | **0.1162***<br>*(0.0089)* |
| **Desktop** x **P2** | — | — | — | -0.0229**<br>*(0.0074)* |
| FE City | Oui | Oui | Oui | Oui |
| FE Date | Oui | Oui | Oui | Oui |

| Within $R^2$ | 0.0065 | 0.0018 | 0.0069 | 0.0140 |
|---|---|---|---|---|
| Observations | 1 635 454 | 1 635 454 | 1 635 454 | 1 620 420 |
| City | 704 | 704 | 704 | 697 |

***Table 2: TWFE estimates national level***

### 5.2. City-Level Mechanism (Orléans)

Six cultural attractions in Orléans allow a sharper test of the physical-presence mechanism: each site's daily gate count is matched to its own Wikipedia page, eliminating city-level aggregation that conflates heterogeneous site types and diffuse online attention. The $\log(1 + x)$ transformation is applied to both the attendance outcome and all pageview regressors.

In the mobile-only specification, the coefficient on $\log(1 + \text{Mobile}_{it})$ is 0.1319 (SE 0.0174, $p < 0.01$), implying that a one-percent increase in mobile pageviews is associated with a 0.132 percent increase in attendance, conditional on all fixed effects and lagged attendance (Table 3, column 1). The desktop-only coefficient is 0.0161 (SE 0.0148) and is not statistically distinguishable from zero (column 2). In the joint specification, the mobile coefficient is essentially unchanged at 0.1314 (SE 0.0175, $p < 0.01$) while the desktop coefficient is 0.0089 (SE 0.0147), also insignificant (column 3). The within-$R^2$ is 0.080 across the mobile specifications, confirming that mobile Wikipedia attention accounts for a meaningful share of residual attendance variance after fixed effects are removed. The site-level pattern mirrors the national results: mobile traffic co-varies with contemporaneous physical presence, while desktop traffic carries no independent signal.

The high city-level heterogeneity establishes that the Wikipedia–tourism relationship varies substantially across destinations. The pattern is consistent with a signal-to-noise interpretation: the mobile coefficient is amplified where Wikipedia pages attract sufficient daily traffic to separate on-site browsing from ambient noise. To probe this, we augmented the Orléans specification by interacting the mobile-pageview coefficient with a binary indicator for above-median Wikipedia page visibility (median: 5.5 daily pageviews). The coefficient for low-visibility venues is 0.084 ($p < 0.01$) ; the implied coefficient for high-visibility venues reaches 0.191, more than twice as large with the interaction term itself significant at the one-percent level. A parallel split based on attendance density yields an analogous gradient (0.059 versus 0.205, both $p < 0.01$). The mechanism therefore operates across the full range of sites but is considerably stronger where page readership is dense enough to carry a reliable signal. The same results have been confirmed with national sample where the association between mobile page view and hotel room night is higher in city with higher page view everything being constant.

| | (1)<br>Mobile | (2)<br>Desktop | (3)<br>Mobile+ Desktop | (4)<br>Lead-lag day 0 |
|---|---|---|---|---|
| log(1 + Mobile_it) | 0.1319***<br>(0.0174) | - | 0.1314***<br>(0.0175) | 0.1355***<br>(0.0191) |

| log(1 + Desktop_it) | - | 0.0161 (0.0148) | 0.0089 (0.0147) | - |
|---|---|---|---|---|
| Y i,t-1 | 0.2576*** (0.0496) | 0.2614*** (0.0495) | 0.2575*** (0.0495) | 0.2414*** (0.0531) |
| Observations | 7,416 | 7,416 | 7,416 | 7,056 |
| R2 | 0.903 | 0.902 | 0.903 | 0.908 |
| Within R2 | 0.080 | 0.070 | 0.080 | 0.088 |
| Venue FE | Yes | Yes | Yes | Yes |
| Date FE | Yes | Yes | Yes | Yes |
| Venue x Year-Month FE | Yes | Yes | Yes | Yes |
| Venue x Day-of-Week FE | Yes | Yes | Yes | Yes |
| Lagged attendance | Yes | Yes | Yes | Yes |
| DK bandwidth | 30 days | 30 days | 30 days | 30 days |

***Table 3: TWFE estimates: Wikipedia pageviews and cultural-site attendance, Orléans***

Same-day timing reduces concerns about reverse causality and slow-moving confounding. The distributed lead-lag model shows a coefficient peaking sharply at $k = 0$ (0.1355, SE 0.0191, $p < 0.01$), falling to 0.0194 at $k = -1$ and 0.0412 at $k = +1$, and indistinguishable from zero beyond ±2 days (Figure 2). Under the Bonferroni simultaneous correction applied to all 61 timing coefficients, $k = 0$ is the only day whose confidence interval excludes zero across the full −30/+30 window. Near-zero pre-event coefficients are inconsistent with anticipatory trends and reduce concerns that common drivers generate both series before the visit date

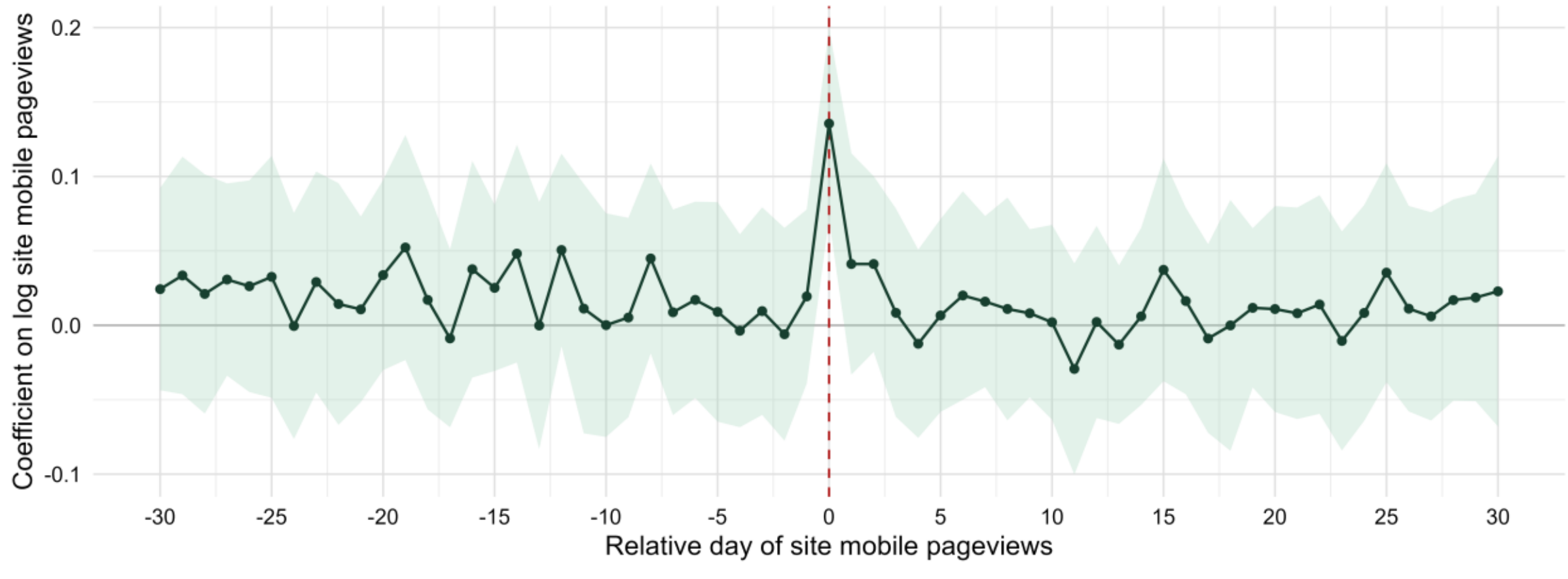

***Figure 2: Timing profile: daily attendance regressed on 61 leads/lags of site-specific mobile pageviews (Orléans, site and date FE, Driscoll-Kraay SE).***

Comparing four prediction models on the 180-day holdout, the joint model M3 (same-day mobile + desktop) achieves the lowest RMSE of 220.4, against 223.1 for the calendar baseline, a reduction of 1.2 percent; out-of-sample correlation rises from 0.462 to 0.499 (Table 4). In six-fold rolling monthly validation, mobile and desktop reduce average RMSE from 207.9 to 202.7 visitors (−2.5%), winning four of six validation months. Desktop alone does not improve the calendar baseline on any metric.

| Model | RMSE | MAE | Log RMSE | Correlation |
|---|---|---|---|---|
| Panel A. Terminal holdout (N = 1,080) | | | | |
| Same-day mobile + desktop | 220.36 | 103.92 | 1.194 | 0.499 |
| Lagged mobile + desktop | 220.55 | 104.11 | 1.202 | 0.490 |
| Same-day mobile only | 222.11 | 104.61 | 1.188 | 0.493 |
| Calendar benchmark | 223.11 | 104.32 | 1.230 | 0.462 |
| Panel B. Rolling monthly validation (6 folds) | | | | |
| Same-day mobile + desktop | 202.74 | 100.07 | 1.181 | 0.559 |
| Same-day mobile only | 204.08 | 100.49 | 1.176 | 0.554 |
| Lagged mobile + desktop | 204.30 | 100.40 | 1.188 | 0.539 |
| Calendar benchmark | 207.87 | 101.10 | 1.209 | 0.516 |

***Table 4: Nowcasting performance in Orléans: terminal holdout and rolling monthly validation.***

# 6. Discussion

## 6.1. Interpretation of Main Findings

The pattern is consistent with the hypothesized distinction between on-the-ground information-seeking (captured by mobile) and at-home trip planning (captured by desktop). Our design identifies contemporaneous co-movement conditional on two-way fixed effects (Angrist & Pischke, 2009) when tourists physically visit a destination, they use smartphones to look up local Wikipedia pages, generating mobile pageviews that are temporally coincident with the tourism activity itself. Desktop pageviews may occur days or weeks before travel during the planning phase, diluting their contemporaneous correlation with realized room-nights.

The magnitude of the mobile elasticity like $(+0.125)$ is modest but economically meaningful when evaluated against the right benchmark. City and date fixed effects absorb more than 90% of room-night variation, leaving only idiosyncratic within-city daily deviations from the seasonal norm. Against this demanding baseline, the finding that a single freely available signal accounts for a statistically robust share of the residual is notable. The attraction-level replication (within $R^2 = 8\%$) confirms that the weaker city-level estimate reflects aggregation dilution and that a micro level mobile page view are informative to estimate touristic activity.

## 6.2. Heterogeneity

The mobile–tourism coefficient varies across destinations following a signal-to-noise logic: where Wikipedia pages attract sufficient daily traffic, on-site browsing is distinguishable from ambient noise, and the effect is considerably stronger. Cities with an almost exclusively conventional hotel supply and a very small leisure-accommodation share face a compounding constraint: the measurement window may miss the relevant demand segment even when Wikipedia visibility is sufficient. The proxy's reliability therefore depends jointly on Wikipedia page visibility and on the alignment between the measurement instrument and the local tourism structure.

# 7. Contributions

We formalize and test a solution to the persistent challenge of interpreting open digital traces for economic measurement (Choi & Varian, 2012; Dellarocas, 2003): device composition (mobile-to-desktop ratio) acts as a behavioural filter separating situated information use from remote browsing.

We demonstrate that Wikipedia serve as a knowledge source but also as a proxy of attention, readership data can serve as a daily-frequency signal for tourism activity. For destination management organizations, it complement real-time indicator of contemporaneous demand. For city planners and public agencies, it is a transparent, auditable input for monitoring disruptions and policy evaluation. For tourism researchers, it is an open, low-cost benchmark that complements or validates proprietary data sources. For Wikipedia's own platform governance, this result show importance of content investment.

## 7.1. Managerial and Policy Implications

For cities with sufficient Wikipedia visibility, real-time mobile pageviews provide a same-day signal of on-the-ground tourism activity. Destination managers can track deviations from seasonal baselines as an early indicator of demand shocks particularly useful during disruptions (transport strikes, extreme weather, public health restrictions) when traditional reporting lags are most costly. The open and replicable nature of the data makes it auditable, a feature absent from proprietary telemetry.

For researchers studying tourism demand, platform economics, or digital trace methodology, mobile Wikipedia data provide an open, zero-cost benchmark. Because the data are non-proprietary and the mechanism is theoretically grounded, result are replicable across time periods, and different cities enabling cumulative knowledge building in a field where proprietary data often constrain external validity.

# 8. Limitations and Future Research

Our tourism demand variable comes from a single hotel chain (Accor), which may not represent all accommodation types. The Accor data primarily capture hotel room-nights, not alternative lodging (vacation rentals, campsites, hostels). This is particularly relevant for interpreting city-level heterogeneity: cities with apparent null or negative effects may not genuinely exhibit a weak Wikipedia–tourism link; rather, the disconnection may arise because Accor's local supply serves a business clientele rather than the leisure tourism that Wikipedia attention captures. Future work with multi-source accommodation data covering hotels, vacation rentals, and campsites could disentangle the proxy's true coverage from this measurement artefact.

Time-varying omitted variables could bias estimates. Local events might simultaneously increase both Wikipedia pageviews and hotel occupancy. The leads-and-lags analysis partially addresses this by showing that the association is concentrated on the day of visit rather than spread across surrounding days, but contemporaneous confounding cannot be fully ruled out without exogenous variation in Wikipedia visibility.

Mobile browsing does not equate to physical presence. Smartphone Wikipedia access also occurs at home, during commutes, or in other non-tourism contexts. The sharp same-day timing profile and the attenuation of the desktop coefficient provide indirect evidence that mobile views track physical mobility, but without fine-grained location data, the physical-presence interpretation remains an inference rather than a direct identification. Future work linking Wikipedia access logs to mobility data would sharpen this identification.

Our analysis operates at the daily level, which is the finest available resolution for both the hotel and Wikipedia data. Within a single day, arrival timing may still cause measurement mismatch: a tourist checking in late afternoon and consulting Wikipedia in the morning could appear as same-day co-movement even though the Wikipedia query slightly precedes check-in. This is unlikely to create large bias but points to a remaining source of imprecision in the physical-presence measurement.

## 9. Conclusion

This paper shows that people reach for their phones and consult Wikipedia when they are physically at a destination and that this behaviour leaves a measurable trace. Using a daily panel of 704 French cities over 2018–2025, we establish a robust positive association between mobile Wikipedia traffic and same-day hotel room-nights, concentrated in the mobile rather than the desktop signal. A micro-validation on six Orléans cultural attractions replicates the pattern with a timing profile sharply concentrated on the day of visit, ruling out reverse causality. These results support a consequential insight: mobile Wikipedia use is a complement to physical presence. Tourists browse Wikipedia while visiting, using it as an on-site information platform to orient, enrich, and deepen their experience of the destination. The relationship is strongest where Wikipedia traffic is substantial enough to carry a reliable signal, and weaker where traffic is too sparse or where the available hotel data do not reflect the relevant tourism segment.

These findings contribute three things. First, they show that the composition of online attention mobile versus desktop contains behaviourally interpretable information beyond aggregate levels, offering a replicable principle for extracting signal from open digital traces. Second, they position Wikipedia as an active information platform whose on-site use by tourists is both an economic signal and a component of the visitor experience: the quality of the platform shapes what travellers can learn at the destination, making content investment a lever for both measurement and visitor satisfaction. Third, mapping boundary conditions, they provide a framework for deciding which actors DMOs, public agencies, researchers, platform contributors can draw the most value from the same open data source, and under what conditions.

### References

Angrist, J. D., & Pischke, J.-S. (2009). *Mostly harmless econometrics: An empiricist's companion*. Princeton University Press.

Bakos, J. Y. (1997). Reducing buyer search costs: Implications for electronic marketplaces. *Management Science, 43*(12), 1676–1692. https://doi.org/10.1287/mnsc.43.12.1676

Bańbura, M., Giannone, D., Modugno, M., & Reichlin, L. (2013). Now-casting and the real-time data flow. In *Handbook of economic forecasting* (Vol. 2, pp. 195–237). Elsevier. https://doi.org/10.1016/B978-0-444-53683-9.00004-9

Bangwayo-Skeete, P. F., & Skeete, R. W. (2015). Can Google data improve the forecasting performance of tourist arrivals? Mixed-data sampling approach. *Tourism Management, 46*, 454–464. https://doi.org/10.1016/j.tourman.2014.07.014

Buhalis, D., & Law, R. (2008). Progress in information technology and tourism management: 20 years on and 10 years after the Internet: The state of eTourism research. *Tourism Management, 29*(4), 609–623. https://doi.org/10.1016/j.tourman.2008.01.005

Chevalier, J. A., & Mayzlin, D. (2006). The effect of word of mouth on sales: Online book reviews. *Journal of Marketing Research, 43*(3), 345–354. https://doi.org/10.1509/jmkr.43.3.345

Choi, H., & Varian, H. (2012). Predicting the present with Google Trends. *Economic Record, 88*, 2–9. https://doi.org/10.1111/j.1475-4932.2012.00809.x

De Los Santos, B., Hortaçsu, A., & Wildenbeest, M. R. (2012). Testing models of consumer search using data on web browsing and purchasing behavior. *American Economic Review, 102*(6), 2955–2980. https://doi.org/10.1257/aer.102.6.2955

Dellarocas, C. (2003). The digitization of word of mouth: Promise and challenges of online feedback mechanisms. *Management Science, 49*(10), 1407–1424. https://doi.org/10.1287/mnsc.49.10.1407.17308

Dickinson, J. E., Hibbert, J. F., & Filimonau, V. (2016). Mobile technology and the tourist experience: (Dis)connection at the campsite. *Tourism Management, 57*, 193–201. https://doi.org/10.1016/j.tourman.2016.06.005

Driscoll, J. C., & Kraay, A. C. (1998). Consistent covariance matrix estimation with spatially dependent panel data. *Review of Economics and Statistics, 80*(4), 549–560. https://doi.org/10.1162/003465398557825

Dunn, O. J. (2012). Confidence intervals for the means of dependent, normally distributed variables. *Journal of the American Statistical Association, 54*(287), 613–621. https://doi.org/10.1080/01621459.1959.10501524

Gao, B., Wang, J., Ding, X., & Guo, Y. (2025). The pitfalls of review solicitation: Evidence from a natural experiment on TripAdvisor. *Management Science, 71*(2), 1671–1691. https://doi.org/10.1287/mnsc.2023.01006

Ghose, A., Goldfarb, A., & Han, S. P. (2013). How is the mobile Internet different? Search costs and local activities. *Information Systems Research, 24*(3), 613–631. https://doi.org/10.1287/isre.1120.0453

Ghose, A., Ipeirotis, P. G., & Li, B. (2012). Designing ranking systems for hotels on travel search engines by mining user-generated and crowdsourced content. *Marketing Science, 31*(3), 493–520. https://doi.org/10.1287/mksc.1110.0700

Giannone, D., Reichlin, L., & Small, D. (2008). Nowcasting: The real-time informational content of macroeconomic data. *Journal of Monetary Economics, 55*(4), 665–676. https://doi.org/10.1016/j.jmoneco.2005.10.021

Godes, D., & Mayzlin, D. (2004). Using online conversations to study word-of-mouth communication. *Marketing Science, 23*(4), 545–560. https://doi.org/10.1287/mksc.1040.0071

Greenstein, S., & Zhu, F. (2016). Open content, Linus' Law, and neutral point of view. *Information Systems Research, 27*(3), 618–635. https://doi.org/10.1287/isre.2016.0643

Greenstein, S., & Zhu, F. (2018). Do experts or crowd-based models produce more bias? Evidence from Encyclopedia Britannica and Wikipedia. *MIS Quarterly*. https://doi.org/10.25300/MISQ/2018/14084

Hinnosaar, M., Hinnosaar, T., Kummer, M., & Slivko, O. (2023). Wikipedia matters. *Journal of Economics & Management Strategy, 32*(3), 657–669. https://doi.org/10.1111/jems.12421

Hollenbeck, B., Moorthy, S., & Proserpio, D. (2019). Advertising strategy in the presence of reviews: An empirical analysis. *Marketing Science, 38*(5), 793–811. https://doi.org/10.1287/mksc.2019.1180

Huertas, A., & Orden-Mejía, M. (2022). Do tourists seek the same information at destinations? Analysis of digital tourist information searches according to different types of tourists. *European Journal of Tourism Research, 32*, 3211–3211. https://doi.org/10.54055/ejtr.v32i.2492

Li, H., Hu, M., & Li, G. (2021). Forecasting tourism demand with multisource big data. *Annals of Tourism Research, 83*, Article 102912. https://doi.org/10.1016/j.annals.2020.102912

Owuor, I., Hochmair, H. H., & Paulus, G. (2023). Use of social media data, online reviews and Wikipedia page views to measure visitation patterns of outdoor attractions. *Journal of Outdoor Recreation and Tourism, 44*, Article 100681. https://doi.org/10.1016/j.jort.2023.100681

Wang, D., Xiang, Z., & Fesenmaier, D. R. (2016). Smartphones in tourism and hospitality marketing: A literature review. *Journal of Travel & Tourism Marketing, 30*(1–2), 178–192. https://doi.org/10.1080/10548408.2014.943458

Wooldridge, J. M. (2010). *Econometric analysis of cross section and panel data* (2nd ed.). MIT Press.

Xiang, Z., Schwartz, Z., Gerdes, J. H., & Uysal, M. (2015). What can big data and text analytics tell us about hotel guest experience and satisfaction? *International Journal of Hospitality Management, 44*, 120–130. https://doi.org/10.1016/j.ijhm.2014.10.013

Yang, Y., Pan, B., & Song, H. (2015). Predicting hotel demand using destination marketing organization's web traffic data. *Journal of Travel Research, 54*(4), 433–447. https://doi.org/10.1177/0047287514544190